\documentclass[aps,prb,showpacs,twocolumn]{revtex4}
\usepackage{graphicx}
\begin{document}

\newcommand{\figureheight}{8.2 cm}
\newcommand{\putfig}[2]{\begin{figure}[h]
        \special{isoscale #1.bmp, \the\hsize \figureheight}
        \vspace{\figureheight}
        \caption{#2}
        \label{fig:#1}
        \end{figure}}

\newcommand{\eqn}[1]{(\ref{#1})}

\newcommand{\be}{\begin{equation}}
\newcommand{\ee}{\end{equation}}
\newcommand{\bea}{\begin{eqnarray}}
\newcommand{\eea}{\end{eqnarray}}
\newcommand{\bean}{\begin{eqnarray*}}
\newcommand{\eean}{\end{eqnarray*}}

\newcommand{\nn}{\nonumber}




\title{Suppression of electron-electron repulsion and superconductivity in Ultra Small Carbon Nanotubes}
\author{S. Bellucci $^1$, M. Cini$^1$ $^2$, P. Onorato $^1$ $^3$ and E. Perfetto$^1$ $^4$ \\}
\address{
        $^1$INFN, Laboratori Nazionali di Frascati,
        P.O. Box 13, 00044 Frascati, Italy. \\
        $^2$Dipartimento di Scienze Fisiche,
        Universit\`{a} di Roma Tor Vergata, Via della Ricerca Scientifica,
  Roma, Italy\\
       $^4$ Instituto de Estructura de la Materia.
        Consejo Superior de Investigaciones Cient{\'\i}ficas.
        Serrano 123, 28006 Madrid. Spain}
\date{\today}
\begin{abstract}

Recently, ultra-small-diameter Single Wall Nano Tubes with
diameter of $ \sim 0.4 nm$ have been produced and  many unusual
properties were observed, such as superconductivity, leading
to a transition temperature $T_c\sim 15^oK$, much larger than that
observed in the bundles of larger diameter tubes.

By a comparison between two different approaches, we discuss the issue whether a superconducting behavior in these carbon
nanotubes  can arise by a purely electronic mechanism. The first approach is based on
the Luttinger Model while the second one, which emphasizes the role
of the lattice and short range interaction, is developed starting
from the Hubbard Hamiltonian. By using the latter model we predict
a transition temperature of the same order of magnitude as the
measured one.

\end{abstract}

\pacs{73.63.Fg, 71.10.Pm, 74.20.Mn, 74.78.Na}

\maketitle

{\bf Introduction}

\

Carbon nanotubes (CNs) are basically rolled up sheets of graphite
(hexagonal networks of carbon atoms)
forming tubes that are only nanometers in diameter and have length up
to some microns. Several experiments in the last 15 years have
shown  their interesting properties\cite{ebbesen}. The nanometric
size of CNs, together with the unique electronic
structure of a graphene sheet, make the electronic properties of
these one-dimensional (1D) structures highly unusual.
In fact, the   electronic properties of CNs
depend on their diameter and chiral angle (helicity) parameterized
by a roll-up (wrapping) vector $(n, m)$\cite{3n}. Hence it follows
that some nanotubes are metallic with high electrical
conductivity, while others are semiconducting with relatively low
band gaps.
CNs may also display different  behaviors
depending on whether they are single-walled carbon nanotubes
(SWNTs, an individual SWNT has typical dimensions: $L\sim1 \mu m$
and $R\sim 1 nm$) or multi-walled carbon nanotubes (MWNTs) that are
typically made of several (typically 10) concentrically arranged
graphene sheets with a radius of about $5 \; {\rm nm}$ and lengths
in the range of $1-100 \; {\rm \mu m}$.

In the following we will study the possibility that a superconducting
behavior can arise, at least in a special class of CNs,  by a purely electronic
mechanism, i.e. neglecting the contribution of phonons, but rather concentrating
on the effect of rescaling  the e-e repulsion for obtaining superconductivity.
Hence, we first review the concept of Luttinger liquid, in particular for CNs,
with the corresponding interaction range and transport behavior, before
describing superconducting correlations in CNs.

\

{\it The concept of  Luttinger liquid - } Electronic correlations
have been predicted to dominate the characteristic features in
quasi one dimensional (1D) interacting electron systems. This
properties, commonly referred to as
 Tomonaga-Luttinger liquid (TLL)
behaviour\cite{TL}, are very different from those of  a Fermi
liquid, because Landau quasiparticles are unstable and the
low-energy excitation is achieved by exciting an infinite number
of plasmons (collective electron-hole pair modes), making the
transport intrinsically different. Thus, the electron-electron (e-e)
interaction modifies significantly the transport properties
(the conductance $G$) also of CNs and leads to the formation of a
Luttinger liquid (LL) with properties very different from those of
the non-interacting Fermi gas\cite{TL,egepj}.

\

{\it Luttinger liquid behavior in carbon nanotubes - } The LL
behaviour implies  the power-law dependence of physical
quantities, such as for the tunneling density of states (DOS), as
a function of the energy or the temperature. The tunneling
conductance $G$ reflects the power law dependence of the DOS in  a
small bias experiment\cite{kf}
 \bea
 G=dI/dV\propto
T^{\alpha_{bulk}} \eea  for $eV_b\ll k_BT$, where $V_b$ is the
bias voltage, $T$ is the temperature and $k_B$ is Boltzmann's
constant. The bulk critical exponent can be obtained in several
different ways and has the form
\begin{equation}\label{al1}
  \alpha_{bulk}=\frac{1}{4} \left(g+\frac{1}{g}-2 \right).
\end{equation}
In  previous papers\cite{noi,npb,prb}, where  we  developed a
Renormalization group (RG) method, in order to study the
low-energy behaviour of the unscreened e-e
interaction in CNs, we obtained \bea\label{K}
 \sqrt{1 + \frac{U_0(q_c)}{ (2
\pi  {v}_F)}} =\frac{1}{g},
\eea
 where $v_F$ is the Fermi
velocity, $ U_0(p) $ corresponds to the Fourier transform of the
1D  e-e interaction potential, and $q_c=2\pi/L$  can be assumed as
the natural infrared cut-off, depending on the longitudinal length
$L$ of the quasi 1D device. Thus, $g$ is a function of the
interaction strength and $g<1$ corresponds to a repulsive
interaction. Thus evidence of LL behavior has been found in many
experiments\cite{ll,ll2,ll3} in SWNT\cite{17}, where a
measurement of  the temperature dependence of the resistance was carried out,
above a crossover temperature $T_c$\cite{Fischer} .

\

{\it Range of the interaction and transport  - } The crucial role
played by  the range of the interaction in CNs, in order to
explain the LL behaviour of large Multi Wall\cite{noi} and
doped\cite{npb,prb} CNs was also analyzed by using RG methods.
Nevertheless at  $T \lesssim 1\; K$ the conductance, $G$, of a
SWNT showed typical Coulomb Blockade (CB) peaks in the zero bias
$G$ and allowed us to investigate the energy levels of interacting
electrons. In this case, crudely described by the CB
mechanism\cite{16}, periodic peaks (Coulomb Oscillations) are
observed in the conductance as a function of the gate
potential\cite{11,10}.

The effects of a long range interaction have to be observed also in
the transport at very low temperature as we discuss below. In a
recent paper\cite{noicb} we investigated the effects of the long
range terms of the interaction in a SWNT and compared our results
with recent experiments at very low temperature $T$\cite{cobden}.
In that paper  we explained the observed damping in the addition
energy for a SWNTs\cite{cobden} at $T\sim 200\, mK$ as an effect
of the long range of the e-e repulsion.

\

{\it Superconductivity - } Experiments have been also carried out to
probe superconducting (SC) correlations in CNs. Clear evidence of
SC correlations was found  in a CN attached to suitable contacts
\cite{7,8}. Supercurrents have been observed in the samples with
SC electrodes reported in Ref.\cite{7}, providing evidence of the
proximity effect in the CN. Moreover, SC transitions have been
measured in nanotube ropes attached to highly transparent
contacts\cite{10b}.

 Recently,
ultra-small-diameter SWNTs (diameter $\sim 0.4\; nm$) have been
produced  inside  zeolite channels (with inner diameter of  $\sim
0.73\; nm$). The ultra small diameter of these tubes gives them
many unusual properties, such as superconductivity, leading
to a transition temperature $T_c \approx 15^o K$\cite{[11]}, much
larger than that observed in bundles of larger diameter tubes
\cite{ropes}.

\

{\it The small diameter SC CN - } In ref.\cite{[11]} the nanotube
diameter $d = 4.2\pm0.2\AA$  is closer to the value calculated for
a $(3,3)$ CN geometry, although the presence of (5,0) nanotubes
cannot be discarded\cite{[14]}. { Next we refer to  these ultra
small nanotubes as US CNs.}
 It has been shown by using the
local-density functional method that the (3,3) nanotubes have the
same band structure of typical armchair nanotubes near the Fermi
level, with a pair of subbands crossing at two opposite
momenta\cite{[14]}. In order to observe the supercurrents, it is necessary
to fabricate thin samples to ensure that there is no potential
barrier within the length of the SWNTs. This is done by further
reducing the length of the CNs to about $50$ to $ 100 nm$, { thus
we assume, in the following, the length of the  US CNs to be $L \sim 50 nm$.}

 \

{\it Electron-phonon assisted superconductivity - } These
experiments have stimulated a significant amount of work at the
theoretical level, in order to understand the origin of the superconducting
transition \cite{9,3,4,5}, where the electron-phonon interaction
 plays a crucial role in the standard superconductivity. In a recent paper\cite{cm}, the
authors verify that the electron-phonon coupling parameter in the
armchair geometry originates mainly from phonons at $q = 2k_F$ and
is strongly enhanced when the diameter decreases.

\

{\it Summary - }In this paper we want to discuss  whether a superconducting
behavior can arise, at least in US CNs,  by a purely electronic
mechanism, i.e. from purely repulsive e-e
interactions. Thus, we do not include phonons in our model even if
we acknowledge that their contribution could be relevant.

 In order to pursue our aim  we investigate the effective range of the e-e
 interaction because the rescaling  of the e-e repulsion
is crucial for obtaining superconductivity. In fact, if we start from the  analysis
of the LL theory for a SWNT developed in
ref.\cite{egepj}, we find   that a purely electronic mechanism
which gives superconductivity needs the screening of the forward
scattering (long range effect), the increasing of the backward
scattering (short range effect) and  relevant effects from the
lattice (very short range effects).

When the short range component of the electron electron
interaction as the effects of the lattice cannot be neglected,
the usual approach to the Luttinger Model breaks down. Therefore,
we have to introduce a model which better describes the very short
range term of the interaction, as well as the localization of the electrons
on the lattice, as it is the case of the Hubbard Hamiltonian.

\

{\bf Luttinger Liquid Approach}

\

In order to analyze the effects due to the size of the CNs on
their properties we first discuss the behaviour of SWNTs with a
radius quite larger than the one of the US CNs.

 The LL theory for a
SWNT was developed in ref.\cite{egepj} where the low-energy theory
 including Coulomb interactions is
derived and analyzed. It describes two fermion chains without
interchain hopping but coupled in a specific way by the
interaction. The strong-coupling properties are studied by
bosonization, and consequences for experiments on  single armchair
nanotubes are discussed. The remarkable electronic properties of
carbon nanotubes are due to the special bandstructure of the $\pi$
electrons in graphite \cite{wallace,divincenzo}. The discussion in
this paper is limited to transport through {metallic} armchair
$(n,n)$ SWNTs, especially we discuss the case of the  (10,10)
CN  with  a length $L=3\mu m$ and we name it  $CN_{10}$. Thus we
have the characteristic dispersion relation of a metallic SWNT
which exhibits two distinct Fermi points at ${\vec K}= (\pm
4\pi/3a,0)$ and $\alpha=\pm$ with a right- and a left-moving
($r=R/L=\pm$) branch around each Fermi point.  These branches are
highly linear with Fermi velocity $v_F\approx 8\times 10^5$ m/s.
The R- and L-movers arise as linear combinations of the $p=\pm$
sublattice states reflecting the two C atoms in the basis of the
honeycomb lattice. The dispersion relation holds for energy scales
$E < D$, with the bandwidth cutoff scale $D\approx \hbar v_F/R$
for tube radius $R$. We choose the $y$-axis points along the tube
direction and the circumferential variable is $0\leq x \leq 2\pi
R$, where $R=\sqrt{3} na/2\pi$ is the tube radius. The lattice
constant is $a=2.46${\AA}.

\

 For what concerns the interaction we distinguish three processes associated with the
Fermi points $\pm K_s$. First, we have ``forward scattering''
($g_2$ with small transferred momentum i.e. $p\sim q_c$). Second,
we have ``backscattering'' ($g_1$ with large transferred momentum
i.e. $p\sim 2 K_s$). Finally, at half-filling there is an
additional ``Umklapp'' process that in our case we neglect, since
the sample is assumed to be doped.

An additional ''Forward scattering'' term  ($f$) which
 measures the difference between intra- and inter-sublattice
interactions, can be introduced following ref.\cite{egepj}. This
term is  due to the hard core of the Coulomb interaction.i.e. it
follows from the unscreened short range component of the
interaction.

\

{\it Electron-electron interaction } -
 Now, by following Egger and Gogolin\cite{egepj}, we introduce the
unscreened Coulomb interaction in two dimensions
 \bea \label{U}
 U_0({\bf r}-{\bf r'})=c_0\frac{e^2}{\sqrt{(x-x')^2+4
R^2 \sin^2(\frac{\varphi-\varphi'}{2})}}. \eea

  Then, we can
calculate $U_0(q)$ as \bea\label{uq}
 U_0(q)\approx \frac{c_0 e^2}{\sqrt{2}}  \left[K_0(\frac{qR}{2})I_0(\frac{qR}{2}) \right],
\label{uq2} \eea where  $K_0(q)$   denotes the modified Bessel
function of the second kind, $I_0(q)$ is  the modified Bessel
function of the first kind and $R$, the CN's radius, acts as a
natural cut off of the interaction. It is clear that the
interaction in eq.(\ref{uq}) does not contain the effects at very
short range due to the  additional forward scattering ($f$
coupling).

 \

{\it The Phase Diagram -} Effective field theory was  solved in
practically exact way by Egger and Gogolin\cite{egepj}. They
obtained for the $CN_{10}$ a value of $g\approx 0.2$ corresponding
to $\alpha_{bulk}\approx 0.32$ in agreement with experiments. They
also predict  the presence of a SC phase due to the effect of
$g_1$ and $f$, but at very low temperatures ($T_b\sim 0.1 m^oK$
and $T_f\lesssim T_b$ see Fig(1)).

 \

Starting from these results a pure electronic mechanism which
gives Superconductivity needs:

i) Screening of the forward scattering, $g_2$  (long range effect
$g>0.5$)

ii) Increasing of the backward scattering, $g_1$ (short range
effect $T_b$)

iii) Relevant effects from the lattice (high value of the
corresponding temperature, $T_f$)

Calculations for a $CN_{10}$  predicts that 1D superconductivity
is the dominant instability only at $T<1m^oK$ with screened
interactions thus a purely electronic mechanism is  not
sufficient. Can this effect be relevant in US CNs?
\begin{figure}
\includegraphics*[width=1.0\linewidth]{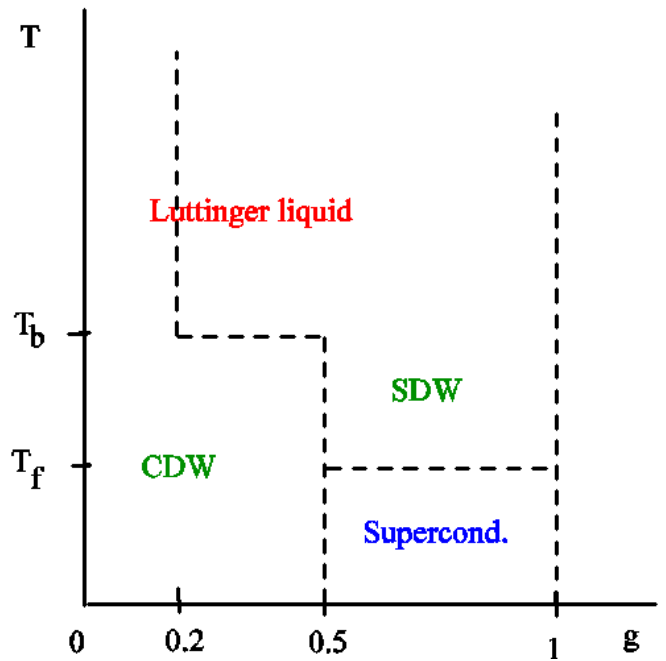}
 \caption{Effective field theory was  solved in a practically exact way by
Egger and Gogolin\cite{egepj}. They concluded  that low
temperature phases matter only for ultrathin tubes or in
sub-mKelvin regime.
    The temperature $T_b$ depends on the strength of the backward scattering term ($g_1$),
    $T_f$ depends on the relevance of the lattice effects while $g$  depends on the forward scattering
    (usually $g_2$) }
\end{figure}

\

{\bf Effects of the screening  }

\

Now we want to discuss some screening effects that can be relevant
in  CNs by focusing on the role played by the size  and on the
reduction of the effective range of the interaction.

\

{\it Intratube screening - } One  electron screening effect can be
taken in account by analyzing how  the interaction dresses  the
bare electron propagator with the polarization. The one-loop
polarizability $\Pi_0(k,\omega_k)$ is given by the sum of
particle-hole contributions within each branch {
\begin{equation}
\Pi_0({ k}, \omega_k) =\frac{1}{h}  \frac{v_F {k}^2}
 { | v_F^2 { k}^2 - \omega_k^2 | }\; \;.
\label{pol0}
\end{equation}
The effective interaction is found by the Dyson equation:
\begin{eqnarray}
U_{eff} ({ k}, \omega_k)  & = & \frac{U_0({ k})}{1+U_0({
k})\Pi_0({ k}, \omega_k)}\label{ueff}.
\end{eqnarray}
This approximation is well justified in 1D as long as we focus on
the long range part of the interaction ($k\sim q_c$)\cite{sol}.
Thus Random Phase Approximation (RPA) for the dielectric function
follows from the previous formula as
$$
\kappa(q,\omega_q)=1+ U_0(q)\Pi_0(q,\omega_q)$$

\

In order to investigate the size depending dielectric function we
can introduce bandwidth cutoff scale $D=v_F \hbar/R$ as the scale
for the UV cut-off energy. Thus we introduce the dimensionless
frequency $\nu= \omega/(v_F q)$ which ranges from $-1/(qR)$ to
$1/(qR)$ and we obtain \bea \nonumber
\kappa(q)&=&1+\int^{1/(qR)}_{-1/(qR)} {d\nu}\left(
U_0(q)\Pi_0(q,\nu)\right)\\ \nonumber &\approx&
1+\frac{U_0(q)}{2\pi \hbar
v_F}\left|\ln\left(\left|\frac{qR+1}{qR-1}\right| \right)\right|.
\eea

Now we can evaluate the size dependent screening  as
$g_i\rightarrow \widetilde{g_i}= g_i/\kappa$, for the different
CNs corresponding to the different processes { (here we name
$\widetilde{g_i}$ the screened interaction)}. For the forward
scattering we obtain a size dependent rescaling of the interaction
\bea
\begin{array}{ccl}
  g_2\rightarrow & \widetilde{g_2}= 0.90 g_2 & \;for\; CN_{10} \\
  g_2\rightarrow & \widetilde{g_2}= 0.61 g_2 &\;for\; US\, CN.
\end{array}
\eea
Concerning the rescaling of $g_1$ the RPA approximation does
not give correct results. In this case we should resort the RG
technique; anyway as shown in Ref.\onlinecite{egepj} $g_1$ gets
modified only at very low temperatures.

\

{\it Screening by the contacts  -} Another source of the electron
screening comes from  the presence of the contacts, and it can be
analyzed by the introduction of two charge images.

As we show in Fig.(2) this kind of screening gives a strong
suppression of the long range component of the interaction (a
constant interaction with infinite range is totally erased) while
the short range one is almost unaffected.
\begin{figure}
\includegraphics*[width=1.0\linewidth]{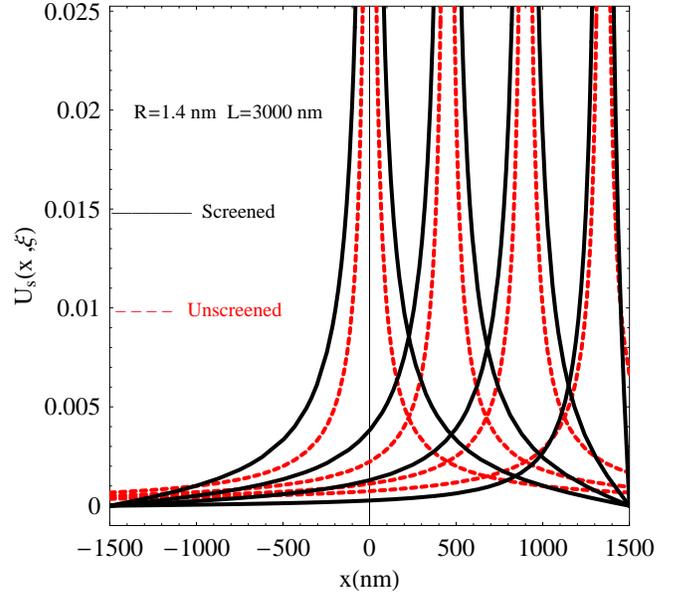}
\caption{{In this figure the  e-e repulsion is shown as a function of
the electron position in a $CN_{10}$. One charge is put at the
center of the CN or at $450,900$ and $1300$ nm from the center.
The screening due to the contacts gives a strong suppression of
the long range component of the interaction (a constant
interaction with infinite range is totally erased) while the short
range one is almost unaffected (increased). }}
\end{figure}
It follows for the forward scattering   a size dependent screening
of the interaction, greater in the US CN than in the usual
$CN_{10}$. We estimate that the long range interaction in the US
CN is reduced by about $7 \%$ more than in the $CN_{10}$, i.e,
$$
\widetilde{g}_2^{USCN} \approx 0.93
\frac{\widetilde{g}_2^{CN_{10}}}{{g}_2^{CN_{10}}}{g}_2^{USCN}
$$

{\it Screening by the zeolite matrix  -} A further important
source of screening arises by the other nanotubes in the
surrounding zeolite matrix. As already pointed out in
Ref.{\onlinecite{gonzperf}}, the intra-tube Coulomb repulsion at
small transfer (i.e. in the forward scattering channel) is
efficiently screened by the presence of electronic currents in
neighbor nanotubes. In the experimental samples of
Ref.\onlinecite{[11]} the carbon nanotubes are arranged in large
arrays with triangular geometry, behaving as a genuine 3D system. By
means of a generalized Random Phase Approximation approach, it is
shown\cite{gonzperf} that the forward scattering parameter $g_{2}$
gets renormalized according to the Dyson equation
\begin{equation}
g_{2} \rightarrow \frac{1}{2\pi v_{F}}\left( \frac{d}{2 \pi}
\right)^{2}\int_{BZ} d^{2}{\bf p} \frac{\phi(k \approx 0,{\bf p})}{1-\Pi (k \approx 0)\phi(k \approx 0,{\bf p})}
 \, ,
\label{screened}
\end{equation}
where $d \approx 1$ nm is the intertube distance in the matrix whose Brillouin zone is denoted by BZ,
 $\Pi(k)=\frac{2}{L}\sum_{q}
\frac{f(\varepsilon_{q+k})-f(\varepsilon_{q})}
        {\varepsilon_{q+k}-\varepsilon_{q}}$, and $\phi(k ,{\bf p})$
is the Fourier of the 3D Coulomb potential with longitudinal momentum $k$ and
2D transverse momentum ${\bf p}$.
The above source of screening provides a large reduction of $g_{2}$ (of a factor $\approx 10^{-2}$), while
the backscattering coupling $g_{1}$ is not affected appreciably\cite{gonzperf}.

\

We conclude this analysis  by pointing out that in the
superconducting samples of Ref.\onlinecite{[11]} the long-range
part of the Coulomb repulsion can be reduced by a total factor of
the order of $\lambda \approx 10^{-3}$, while the short-ranged one
is essentially unmodified, at least in the temperature range of
the experimental conditions.

\

{\it Unscreened parameters: the short range component - } As we
discussed above the short range interaction contributes to two
fundamental parameters. The first one, the backward scattering
term,  has to be stronger in small diameter CN, in fact we
calculate $g_1\sim 0.067 (2\pi v_F)$ in  $CN_{10}$ ($g_1/g_2\sim
0.003$) and $g_1\sim 0.45 (2\pi v_F)$ in  US CN ($g_1/g_2\sim
0.04$) thus
 $g_1^{US} \sim 6.5g_1^{CN_{10}}$.

The temperature $T_b$ reported in the phase diagram of Fig.(1) was
calculated in ref\cite{egepj} as
$$
k T_B\propto D e^{\frac{2\pi v_F}{g_1}},
$$
and was estimated for the $CN_{10}$ in the order of $T_b\sim 0.1
m^oK$. It follows that  in US CNs $T_b$ should be several orders
of magnitude  larger than the one predicted for a $CN_{10}$ with a
factor compatible with the observed critical temperature.

\

The coupling constant $f>0$, even though it can be assumed as a forward
scattering term, strongly depends on the nanotube geometry and
the short range component of the interaction. In fact when we
consider two interacting electrons at a very short distance, the fact of
belonging to the same sublattice, or not, becomes relevant as we
show in Fig.(3).
\begin{figure}
\includegraphics*[width=1.0\linewidth]{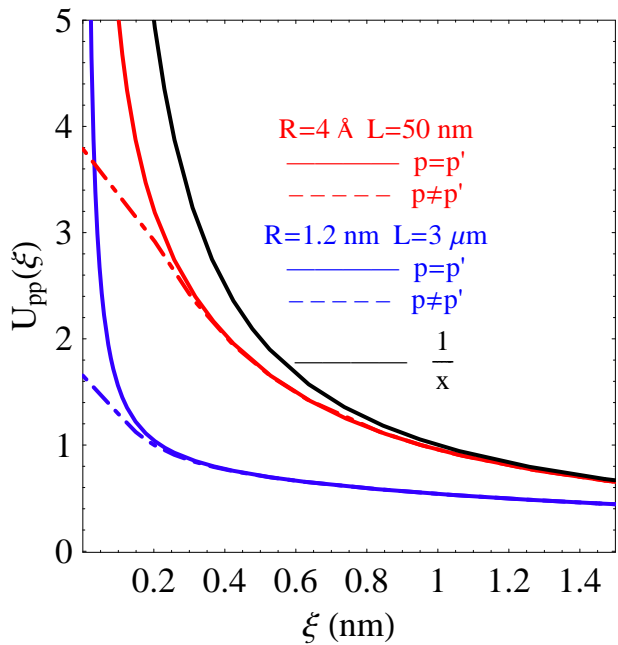}
\caption{{In figure the  e-e repulsion is shown. The interaction
between electrons belonging to different sublattices $p\neq p'$
has a short range component quite smaller than the interaction
between electrons belonging to the same sublattice $p=p'$.}}
\end{figure}
 In order to calculate
$f$ on the wrapped graphite lattice, we can start from the
microscopic arrangement of carbon atoms around the waist of the
armchair SWNT. Following ref.\cite{egepj} is quite easy to
demonstrate that $f\propto \frac{1}{n}$ for a $(n,n)$ CN, so that
we obtain that is $10/3$ times larger in the US CN than in the
$CN_{10}$.

\

{\it Breakdown of the Luttinger approach -} Now we analyze the
additional forward scattering $f$. It corresponds to\cite{egepj}
$\delta V_p= U_{++}-U{+-}$, where $U_{p,p'}$ is the interaction
between electrons belonging to different sublattices ($p,p'$). In
Fig.(3) we plot  $U_{p,p'}$ both for the $CN_{10}$ and US CN. We
observe that, because of the rapidly oscillating phase factor, the
only non-vanishing contribution to $g_1$ comes from $|x-x'|\leq
a$\cite{egepj}. Therefore   we can conclude that a local
interaction is relevant in CNs. This very short range interaction
is strongly suppressed at a distance  much larger than $\ell\sim
0.3 nm$
 enforcing  the validity of a
short range model (like the Hubbard one)   for the small radius CN
where this contribution, $f$, is comparable to $g_2$. We suppose
that the LL theory, which just predicts a Charge Density Wave
instability for $g\lesssim 0.5$, could not include the strong
effects of the lattice and short range component of the interaction
which could be dominant in US CNs. In fact, because of the
dominance of the the short range component of the electron
electron interaction, the effects of the lattice in the US CN
cannot be neglected or treated as a perturbation.

 Thus we suggest the presence of a SC phase, where
the lattice effects and the very short range interaction become
dominant. In our opinion this kind of system should be better
described in the Hubbard-like  approach than by using the LL
theory.

\

{\bf Hubbard Model }

\

{ The possibility of a superconducting phase in CNs  in the
framework of the Hubbard model was discussed in some papers in the
past \cite{psc,krotov}.

\

Krotov and coworkers \cite{krotov} used an on-site ($U$) and
nearest-neighbor interaction ($V$) to model the screened e-e
repulsion. It follows, for slightly doped samples, that a
superconducting phase is present for values of parameters
$\frac{V}{U}<0.55$, signaling that the pure Hubbard model ($V=0$) would show superconductivity.
 Our previous discussions about the screening
of the interaction and the corresponding reduction of its range,
confirm that the superconductivity phase is supported by the size
of the US CN. Unfortunately  the approach of
Ref.\onlinecite{krotov} does not allow the estimation of the
critical temperature $T_c$.

\

In Ref.\cite{psc,cini} the authors explored an electronic mechanism
which {\em per se} leads to bound pairs starting from the pure
Hubbard model.  The notion that pairing can arise by a purely
electronic mechanism, i.e. from purely repulsive e-e
interactions, was put forth by Kohn and Luttinger long
ago\cite{kohn}. They suggested that for large odd values of the
relative angular momentum two electrons could stay enough far
apart from each other to take advantage of the Friedel
oscillations of the screened Coulomb potential. In the approach
based on a 2$d$ Hubbard model the first-order Coulomb repulsion is
removed by symmetry.

In Ref.\onlinecite{psc} it was shown that the Hubbard Hamiltonian
for a CN admits two-body singlet eigenstates with no double
occupancy, called
  $W=0$ pairs. The electrons forming a  $W=0$ pair have no direct
interaction and are the main candidates to achieve bound states in
purely repulsive Hubbard models already used for the Cuprates
\cite{cini}.

Following the approach developed in Ref.\onlinecite{psc} we can
evaluate  the superconducting gap, ${ \Delta}$, which is strongly dependent
on the size of the CN. In fact, as discussed in
Ref.\onlinecite{egepj} and refs. therein~\cite{krotov,balents}, in
the language of the Hubbard-like models we have $f\propto U$, thus
the on-site  Coulomb interaction $U$ can be assumed proportional
to $1/n$. Thus the energy gap $\Delta$ can be estimated about $3$
orders of magnitude greater in the US CN than in the $CN_{10}$. In
the US CN away from half-filling we can evaluate the values of
$\Delta$. The BCS theory estimates the zero-temperature energy gap
$$\Delta(0)\approx 1.76 k_B T_c,$$
thus for a US CN we are able to estimate the crossover temperature
$$
T_c\approx 5 \div 50 ^oK
$$
of the same order of the measured one while the corresponding
$T_c$ for the $CN_{10}$ is of the order of the $m^-K$ in agreement
with the discussed predictions of Ref.\onlinecite{egepj}.

\

Now we want to discuss the consistency of the above results by
using two different criteria. In order to do that it is useful to
estimate the  coherence length $\xi_c$ of the Cooper pairs in US
CNs which can be obtained by the well known relation
$$
\xi_c=\frac{\hbar v_F}{\pi \Delta(0)}.
$$
We estimate $\xi_c $ of some tens of $nm$s, which is of the same order of
the CN's length.

The first criterion in order to establish the validity of the
Hubbard model concerns the value of the interaction at distance
$r\sim \xi_c$ which  has to be many times smaller than the energy
gap $\Delta$,
$$
\frac{\lambda e^2}{\xi_c}\ll \Delta(0).
$$
In our case this condition is verified ($\lambda\sim 10^{-3}$) because of
the strong screening, especially due to the zeolite matrix.

The second criterion in order to ensure that the electrons forming
the Cooper pairs feel a short range interaction is based on the
comparison between $\xi_c$ and the characteristic range of the
screened interaction $\ell$ (see Fig.(3)), $$ \xi_c \gg \ell.$$
Also this condition is fairly fulfilled and confirms the
consistency of the approach based on the Hubbard model. }

{\bf Conclusion } The discovery of superconductivity at 15 $^{0}K$
in SWNT challenges the usual phonon mechanism of superconductivity
 no less than the similar discovery of superconductivity \cite{Weller} at 6.5
 degrees in C$_{6}$Yb and at 11.5 degrees in
 C$_{6}$Ca Intercalated Graphite. In both cases the possibility that
  the electrons themselves  provide the driving force must be taken
  into account. This means that it should be possible in some
  fashion to go all the way from the e-e repulsion to
  an effective attraction and to bound pairs. This is a time-honored
  dream that potentially has conceptual appeal as well as important practical
  implications, but a convincing mechanism must predict which
  properties of the material are important to produce
  superconductivity. For instance, one could consider replacing
  phonons by plasmons in the usual mechanism; the weakness of such
  an approach is that plasmons exist in any material, and have
   comparable frequencies,  so everything
  should superconduct at several $^{0}K$. The $W=0$ mechanism  is based on the point
  symmetry of the lattice and predicts pairing in Cuprates and in
  Graphite-based materials on the same ground.
   Unlike the original
  suggestion by Kohn and Luttinger\cite{kohn}, which has not yet
  been borne out by experiments, SWNT  (in vacuo or in a matrix) and
  intercalated Graphite are anisotropic inhomogeneous systems. They
  are so different that the respective mechanisms can differ in important
  ways, yet they have the honeycomb lattice in common, i.e. the local symmetry is the same
  and leads to $W=0$ pairs. The hexagonal
  lattice leads to pairing in a Hubbard model, but the long-range
  part of the
  repulsion must be disposed of, if we want the on-site repulsion to
  operate undisturbed, and here nanotubes and Graphite obviously
  pose quite different problems;  SWNT  is the harder case to
  understand, since screening works better in 3d than in 1d. In this
  paper we have focused  on  this problem and provided possible
  explanations based on the residual screening effects augmented by
  a powerful matrix and contacts contribution. Furthermore, we pointed
  out that the range of the interaction must be small compared to
  the pair size, but
  it should not necessarily be cut
  down to a lattice  parameter. Although this proposal clearly needs
  further scrutiny before being validated, we feel it is serious
  enough to warrant further investigation.


\bibliographystyle{prsty} 

\bibliography{}

\end{document}